\documentclass[prl,twocolumn,amsmath,amssymb,groupedaddress,superscriptaddress]{revtex4}
\usepackage{graphics}

\def\be{\begin{equation}}
  \def\ee{\end{equation}}

\begin{document}

\title{Effective quantum spin systems with trapped ions}

\author{D. \surname{Porras}}
\email{Diego.Porras@mpq.mpg.de} \affiliation{Max-Planck-Institut
f\"ur Quantenoptik,
  Hans-Kopfermann-Str. 1, Garching, D-85748, Germany.}
\author{J.~I. \surname{Cirac}}
\email{Ignacio.Cirac@mpq.mpg.de} 
\affiliation{Max-Planck-Institut f\"ur Quantenoptik, Hans-Kopfermann-Str. 1, Garching, D-85748, Germany.}

\begin{abstract}
We show that the physical system consisting of trapped ions interacting with lasers may undergo a rich variety of quantum phase transitions.
By changing the laser intensities and polarizations the dynamics of the internal states of the ions can be controlled, in such
a way that an Ising or Heisenberg-like interaction is induced between effective spins.
Our scheme allows us to build an analogue quantum simulator of spin systems with trapped ions, and observe and analyze quantum phase transitions with unprecedent opportunities for the measurement and manipulation of spins.

\end{abstract}

\date{\today}

\maketitle

{\em Introduction.}
Quantum spin models are a paradigm for the study of many-body effects, and show intriguing phenomena like the existence of quantum phase transitions at certain values of the parameters that govern the spin Hamiltonian \cite{Sachdev}.
In this work we show that quantum phase transitions can be induced by lasers in ion traps, where the internal states of the ions play the role of effective spins. In our scheme an Ising, XY, or XYZ spin-spin interaction is transmitted by collective vibrational modes and can be switched and tuned by the lasers and by the choice of trapping conditions. The understanding of quantum spin models has been hindered by the fact that they are very hard to be numerically simulated. This work presents a way to circumvent this difficulty by using a system of trapped ions as an analog quantum simulator for magnetic systems \cite{Feynman,ionsimulator}. 
Other schemes for the simulation of quantum problems \cite{Lloyd,NielsenChuang,Lidar} rely on the use of a quantum computer \cite{CiracZoller95,Wineland95,Wineland2003,BlattCZgate,Wunderlich}, or on the stroboscopic change of quantum parameters \cite{Sorensen,Jane}. 
Our proposal is not based on quantum gates, so that the requirements for its implementation are much less stringent, and it is not exposed to the errors accumulated in stroboscopic methods. 

The main advantage
of ions is that they can be trapped and cooled very efficiently,
and therefore they can be stored at fixed positions in space \cite{note}.
Furthermore, their internal states can be precisely manipulated
using lasers, and measured with basically 100\% efficiency \cite{Ionreview}.
Our proposal, thus, would make feasible to study quantum phase transitions 
in this particular experimental system, where one is able to tune the range and strength of the interaction, and manipulate and measure single effective spins, something that is not possible in solid-state magnetic systems.  

\begin{figure}
  \centering
  \resizebox{3.4in}{!}{
    \includegraphics{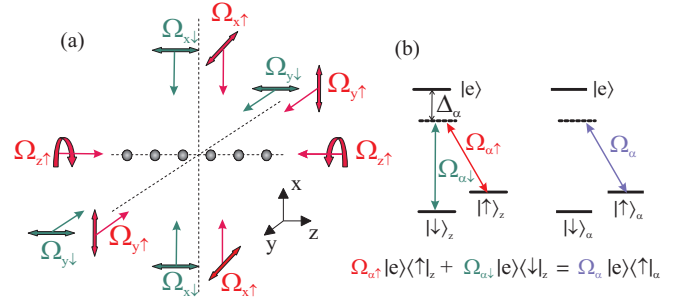}}
  \caption{
  (color online) Possible laser configuration such that the ions
  experience a force along the $\alpha$ = $x$, $y$, and $z$ directions, if
  they are in the state $| \! \uparrow \rangle_{\alpha}$ (Fig (a)).
  Each pair of copropagating lasers independently excite a Raman
  transition $| \! \uparrow \rangle \to |\!\downarrow \rangle$ with different detunings
  from the virtual excited state $|  e \rangle $ (such that they do not interfere).
  The relative phase of each pair of copropagating lasers along the
  direction $\alpha$ is chosen such that the state $|\!\downarrow \rangle_\alpha$
  is dark. Thus, they effectively interact only with the level
  $|\!\uparrow \rangle_\alpha$ [see Fig (b)]. 
  Counterpropagating lasers form a
  standing wave pattern which pushes the ions along the direction
  $\alpha$ only if they are in state $|\!\uparrow\rangle_\alpha$. Note that along the
  $z$ direction only one laser is needed. Of course, this direction does
  not need to coincide with the trap axis for the case of a linear ion trap.}
\label{fig1}
\end{figure}

{\em Spin-spin interaction.}
Our scheme works with a set of $N$ ions trapped by electric and/or magnetic
forces. For example, they could be ions in a linear Paul trap, in
microtraps in 1 or 2 dimensions, or in a Penning trap. 
In order to simplify the discussion that follows, we focus here on the one dimensional case: a chain of trapped ions. Let us assign the $z$ ($x$, $y$) vectors to the axial (radial) directions relative to the chain. 
We will assume that the ions have two internal ground hyperfine levels,
which play the role of components of an effective spin 1/2 (our
scheme can be easily extended to simulate larger spins). We denote
by $\sigma^{\alpha}$, $\alpha = x,y,z$, the Pauli operators, and by
$| \! \downarrow \rangle_{\alpha}$ and $| \! \uparrow \rangle_{\alpha}$ the corresponding
eigenstates. All the ions are driven by the same off resonant
laser beams propagating along the three spatial directions 
(Fig. \ref{fig1}). In the most general configuration, we will
assume that the lasers propagating along $\alpha$ push the
ions in that direction provided they are in states
$|\! \uparrow \rangle_\alpha$. This can be achieved by
choosing the relative phases of the lasers so that the state
$|\! \downarrow \rangle_\alpha$ is dark with respect to the lasers
propagating along the direction $\alpha$ (see, e.g., Fig.\
\ref{fig1}).

As a result of Coulomb repulsion, vibrational degrees of freedom are collective modes (phonons):
$H_v = \sum_{\alpha,n} \hbar \omega_{\alpha,n} a^{\dagger}_{\alpha,n} a_{\alpha,n}$, 
where $a^{\dagger}_{\alpha,n}$  ($\hbar \omega_{\alpha,n}$), 
are the operator (energy) of the $n$ mode in the $\alpha$ direction. Our Hamiltonian includes also the force produced by the lasers ($H_f$), and the action  of effective magnetic fields ($H_m$):
\begin{eqnarray}
H   &=&   H_v + H_f + H_m .
\nonumber \\
H_f &=& - 2 \sum_{\alpha,i} \! \!  F_{\alpha} q_{\alpha,i} | \! \uparrow \rangle \langle \uparrow \! |_{\alpha,i} , \hspace{0.2cm}
H_m = \sum_{\alpha,i} B^{\alpha} \sigma^{\alpha}_i.
\label{hamiltonian}
\end{eqnarray}
The fields $B^{\alpha}$ can be simulated by lasers acting on the internal transition of the ions only.
$q_{\alpha,i}$ are the three spacial coordinates of each ion, and $i$ refers to the site in the chain. The local coordinates can be expressed in terms of the collective modes:
\begin{equation}
H_{f} = - \sum_{\alpha, i, n} 
F_{\alpha} \frac{{\cal M}^{\alpha}_{i,n}}{\sqrt{2 m \omega_{\alpha,n} / \hbar}} 
\left( a^{\dagger}_{\alpha,n} + a_{\alpha,n} \right) \left(1 + \sigma_i^{\alpha} \right) .
\label{conditional.force}
\end{equation}

A suitable method to get the promised spin Hamiltonian is to apply the following canonical transformation:
\begin{eqnarray}
U = e^{-{\cal S}}. &&\hspace{0.2cm} {\cal S} = \sum_{\alpha,i,n} \eta^{\alpha}_{i,n} \left(1 + \sigma^{\alpha}_i \right)
 \left( a^{\dagger}_{\alpha,n} -  a_{\alpha,n} \right) .
\nonumber \\
{\eta}^{\alpha}_{i,n} &=& F_{\alpha} \ \frac{{\cal M}^{\alpha}_{i,n}}{\hbar \omega_{\alpha,n}} 
\sqrt{\frac{\hbar}{2 m \omega_{\alpha,n}}} .
\label{canonical.transformation}
\end{eqnarray}
In the new basis the Hamiltonian (\ref{hamiltonian}) includes an effective spin-spin interaction:
\begin{equation}
e^{- {\cal S}} H e^{\cal S} = H_v + \frac{1}{2} \sum_{\alpha,i,j} J^{\alpha}_{i,j} \sigma_i^{\alpha} \sigma_j^{\alpha} + \sum_{\alpha, i} 
{B'}^{\alpha} \sigma_i^{\alpha} + H_E,
\label{transformed.hamiltonian}
\end{equation}
where
\begin{equation}
- J^{\alpha}_{i,j} \! = \! 
\sum_n \frac{F_{\alpha}^2}{m \omega^2_{\alpha,n}} 
{\cal M}^{\alpha}_{i,n} {\cal M}^{\alpha}_{j,n} = 2 \sum_n \eta^{\alpha}_{i,n} \eta^{\alpha}_{j,n}  \hbar \omega_{\alpha,n} .
\label{effective.interaction.1}
\end{equation}
The effective magnetic fields receive a contribution from the pushing forces ${B'}^{\alpha} = B^{\alpha} + F_{\alpha}^2/(m \omega_{\alpha}^2)$, where $\omega_{\alpha}$ are the trapping frequencies in each direction. Note that the extra term in ${B'}^{\alpha}$ does not depend on the site of the ion. $H_E$ describes the residual coupling between the effective spins and the collective vibrational modes.
To lowest order in $\eta^{\alpha}_{i,j}$, it is given by the following expression:
\begin{eqnarray}
H_E &=&  - \frac{1}{2} \underset{i,n,m}{ 
\underset{\alpha,\alpha'}{\sum} } 
\eta^{\alpha}_{i,n} \eta^{\alpha'}_{i,m} \hbar \omega_{\alpha, n}
\nonumber \\
& & ( a^{\dagger}_{\alpha,n} + a_{\alpha,n} ) ( a^{\dagger}_{\alpha',m} - a_{\alpha',m} ) 
\left[ \sigma_i^{\alpha}, \sigma_i^{\alpha'}  \right] .
\label{error.heisenberg}
\end{eqnarray}
In the transformed basis, $H_{f}$ disappears, and we get an anisotropic Heisenberg (XYZ) interaction between effective spins.
$H_E$ is a perturbation that limits the accuracy of the simulation, and can be neglected if the adimensional coefficients $\eta^{\alpha}_{i,n}$ are small enough. We will discuss later this approximation and consider now the properties of the effective spin-spin interaction.

$J^{\alpha}_{i,j}$ depends on the characteristics of the vibrational modes in the corresponding $\alpha$ direction. 
The ions occupy the equilibrium positions $z_i^{0}$ along the chain, in such a way that the Coulomb repulsion is balanced with the trapping forces.
The second derivatives of the Coulomb energy with respect to these displacements determine the elastic constants of the chain \cite{Steane97,Dubin}:
\begin{eqnarray}
V &=& \frac{1}{2} m \sum_{\alpha,i,j} {\cal K}^{\alpha}_{i,j} {q}^{\alpha}_{i} {q}^{\alpha}_{j} . \nonumber \\
{\cal K}^{\alpha}_{i,j} 
&=& \left\{ \begin{array}{ll} \omega^2_{\alpha} - c_{\alpha} \sum_{j' (\neq i)} \frac{e^2 / m }{|z_i^{0} - z^{0}_{j'}|^3} \hspace{0.5cm} i=j
                 \\  \hspace{0.5cm} + c_{\alpha} \frac{e^2 / m }{|z_i^{0} - z_j^{0}|^3} \hspace{0.5cm} i \neq j \end{array} \right. ,
\label{motion}
\end{eqnarray}
where $c_{x,y} = 1$, $c_z = - 2$. The unitary matrices $\cal M^{\alpha}$ in (\ref{conditional.force}) diagonalize the vibrational hamiltonian:
${\cal M}^{\alpha}_{i,n}{\cal K}^{\alpha}_{i,j} {\cal M}^{\alpha}_{j,m} = \omega_{\alpha,n}^2 \delta_{n,m}$, and Eq. (\ref{effective.interaction.1}) implies that we can express the effective interaction in the following way:
\begin{equation}
J^{\alpha}_{i,j} = - \frac{F_{\alpha}^2}{m} \left(1/{\cal K}^{\alpha} \right)_{i,j}  .
\label{effective.interaction.2}
\end{equation}
If we assume a constant distance between ions, $d_0$, then
we can summarize the properties of the vibrational modes by means of the parameters
$\beta_{\alpha} = |c_{\alpha}| e^2 / m \omega_{\alpha}^2 d_0^3 $. We distinguish two cases: 
{\em (i)} Stiff modes: $\beta_{\alpha} \ll 1$  implies that the Coulomb interaction can be considered as a perturbation to the trapping potential, so that the inverse of the elasticity matrix ${\cal K^{\alpha}}$ can be calculated to first order in the non-diagonal terms in Eq. (\ref{motion}): 
$J^{\alpha}_{i \neq j} = c_{\alpha} F_{\alpha}^2 e^2/(m \omega_{\alpha}^2)^2|z_i^{0} - z_j^{0}|^3$. 
In this case we get a spin-spin interaction with a dipolar decay law, that is antiferromagnetic (ferromagnetic), 
if it is transmitted by the radial (axial) vibrational modes.
{\em(ii)} Soft modes: $\beta_{\alpha} \gg 1$ implies that the Coulomb interaction is important against the trapping of the ions in the $\alpha$ direction,
a situation that is discussed below in connection with concrete trapping conditions.

{\em Experimental realizations.}
We discuss now the experimental setups that can be used to implement the simulation of the quantum spin hamiltonians exposed above.

{\em (i) Arrays of ion microtraps.}
In this case ions are confined individually by microtraps, in such a way that 
they form a regular array in 1 or 2 dimensions \cite{Cirac00}. A 1D array of microtraps can be realized, for example, by using a linear Paul trap,
and an additional off-resonant standing-wave that creates a periodic confining potential in the axial direction.
In the 1D case, forces in the three spacial directions would simulate an effective XYZ interaction, antiferromagnetic, or ferromagnetic, for the ($x$, $y$), or $z$ spin components, respectively.
If the radial trapping frequencies $\omega_{x,y}$ are large enough, the stiff limit could be reached, and the spin-spin interaction would have a dipolar decay. By relaxation of the radial trapping frequencies the spin-spin interaction could be controlled and, in particular, the interaction between second neighbors could be suppressed, similarly to the case of linear Paul traps (see below). The effective XYZ interaction, with the addition of effective magnetic fields would allow to study a rich variety of quantum phase transitions.

\begin{figure}
  \centering
  \resizebox{1.6in}{!}{%
    \includegraphics{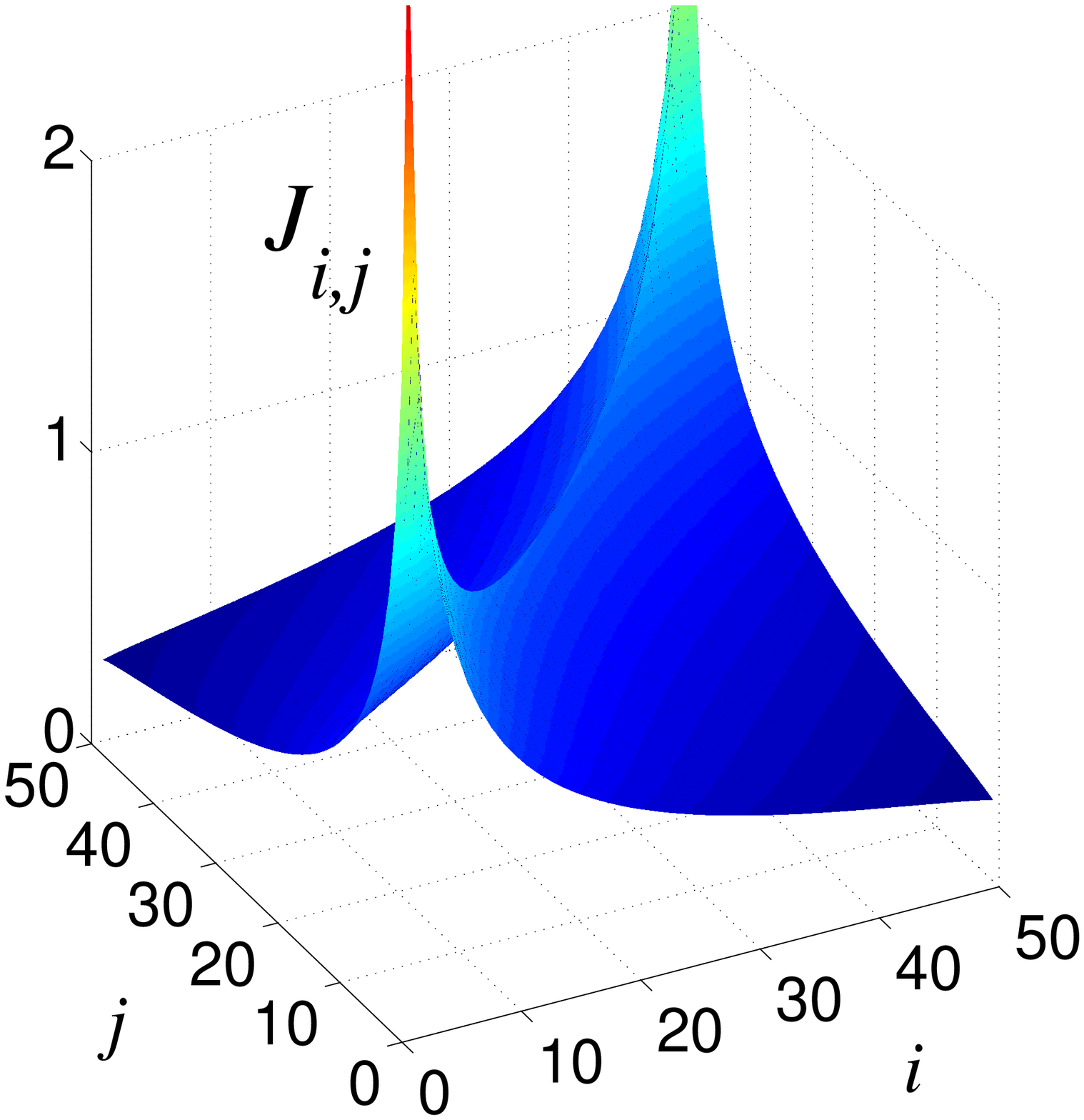}}
  \resizebox{1.6in}{!}{%
    \includegraphics{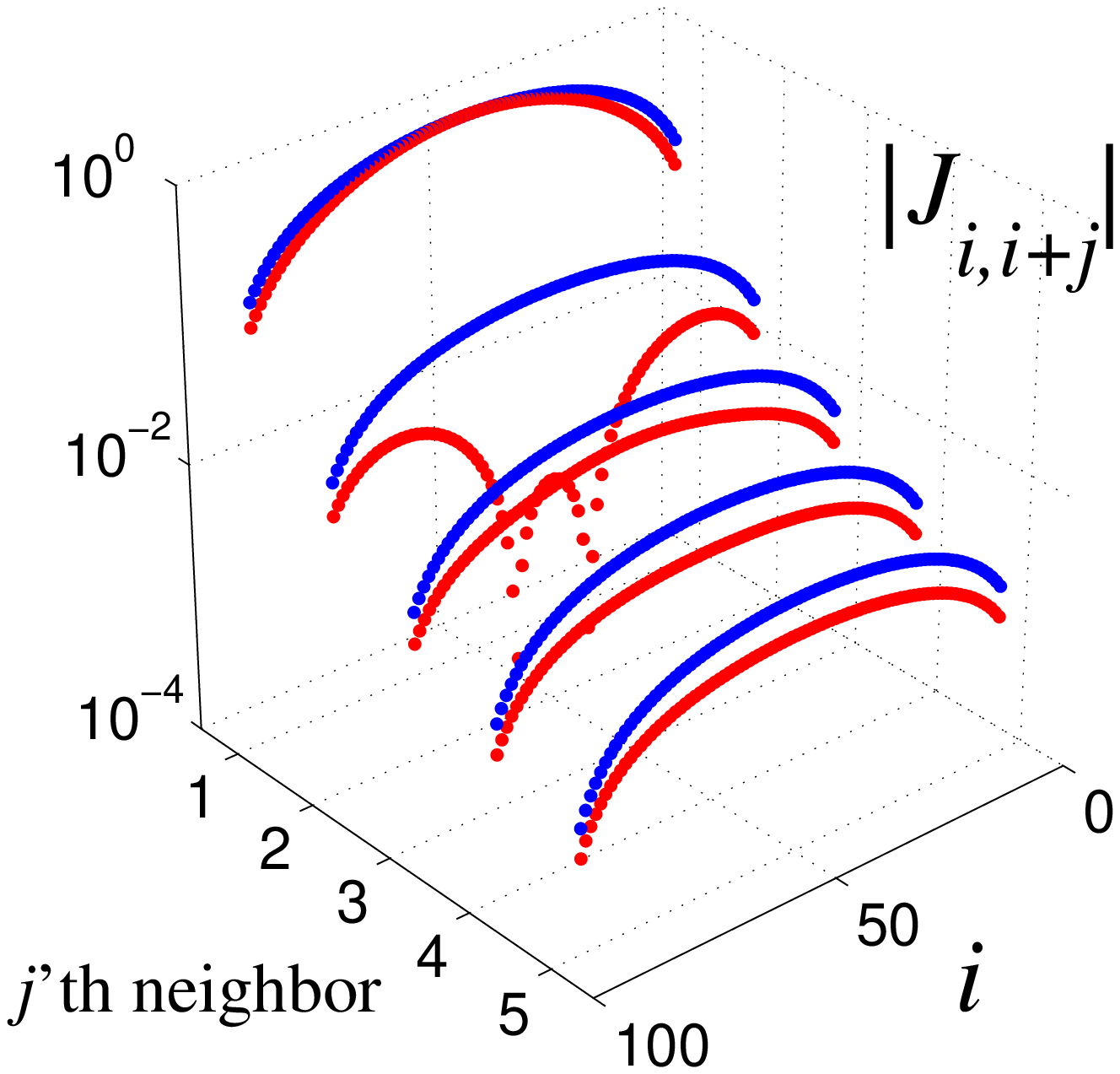}}
  \caption{(color online) Effective spin-spin interactions in a Coulomb chain. Left: long-range ferromagnetic interaction transmitted by the longitudinal modes ($N =$ 50 ions). Right: short-range antiferromagnetic interaction transmitted by the radial modes ($N =$ 100). For $\beta_{x,y}$ = 0.01 (lower curves), the effective interaction shows a dipolar decay: $J_{i,j} \propto |i-j|^{-3}$, while for $\beta_{x,y}$ = 0.1 (upper curves), the second-nearest neighbor interaction is suppressed.}
\label{fig2}
\end{figure}

{\em (ii) Linear Paul traps.}
Paul traps are currently used to cool and arrange ions in 1D Coulomb crystals, in which the distance between ions is not constant.
We can, however, define an averaged lattice constant $d_0$, in order to understand the qualitative properties of the vibrational modes. We distinguish two cases, depending on the orientation of the pushing forces relative to the chain:
{\em (a) Axial force.} The equilibrium position of the ions are such that $\beta_z \gg 1$ \cite{Dubin}.
The center of mass mode has the main contribution in Eq. (\ref{effective.interaction.1}), so that we get a long-range ferromagnetic interaction. 
Our numerical calculation for a chain of $N$ = 50 ions (Fig. \ref{fig2} (left)), shows that a force in the $z$ direction results in an interaction, $J_{i,j}$, with a range comparable to the size of the chain.
{\em (b) Radial force.} The characteristics of the effective spin-spin interaction drastically change when lasers push the ions in the radial ($x$, $y$) directions. The stability of the chain against zig-zag deformation implies that $\beta_{x,y} < 1$ \cite{Dubin}. If the trapping frequencies $\omega_x$, $\omega_y$ are large enough, the limit in which the spin-spin interaction shows a dipolar decay can be easily reached.
%
%
For $\beta_{x,y} \geq 0.05 $, the spin-spin interaction departs from the dipolar case, and the main contribution comes now from the collective stretched mode. With the proper choice of trapping frequencies, the second-neighbor term in the dipole interaction can be suppressed (see Fig. \ref{fig2} (right)). The possible experimental configurations in a linear Paul trap are the same as those for a linear array of ion microtraps, with the difference that the modes in the axial direction transmit a long-range interaction.  

{\em (iii) Penning traps.}
Finally, 2D Coulomb crystals could be used for the simulation of 2D quantum spin systems.
Cooling of ions in a 2D hexagonal lattice has been demonstrated in Penning traps \cite{Itano,Mitchell}. This experimental setup has the additional advantage that the hexagonal lattice could allow to simulate the effect of magnetic frustration. For example, a pushing laser in the direction perpendicular to the 2D crystal would induce an antiferromagnetic Ising interaction. The decay of this interaction would also be dipolar, if the 2D vibrational modes transverse to the crystal plane are in the stiff limit.

In all the three experimental systems discussed above the simplest laser configuration would consist of only one force in a given direction, so that an Ising interaction $\propto J \sigma^z \sigma^z$ would be simulated. 
With the addition of an effective transverse field $\propto B \sigma^x$, the spin system would present a quantum phase transition between a ferromagnetic (or antiferromagnetic) ordered ground state ($J \gg B^x$), and a disordered ground state ($B^x \gg J$), at a value $J \approx B^x$ \cite{Sachdev}.
For example, one could proceed as follows: (1) Prepare the state $| \downarrow \ldots \downarrow \rangle$. (2) Switch on adiabatically the magnetic field $B^x$. (3) Switch on adiabatically the effective interaction up to a given value $J$. (4) Measure the global fluorescence. This sequence could be repeated with several values of $J$, starting with $J \ll B^x$, and increasing the value of the spin-spin interaction in each experiment.
The phase transition would be evidenced in the global fluorescence, that would show the population of the internal states of the ions, and, thus, the emergence of ferromagnetic (or antiferromagnetic) order in the effective spins. Furthermore, the possibility of addressing single ions could allow us to manipulate the many-body ground state, create excitations (spin-waves), or study the entanglement between effective spins. The non-equilibrium quantum dynamics of the Ising model could also be studied, for example, by switching abruptly the interaction on.

{\em Validity of the effective spin Hamiltonian.}
Assume that the ions are initially in the internal state $| \psi_i \rangle$, that would evolve to $|\psi_f \rangle = exp(- i H_S t/\hbar) |\psi_i \rangle$, under the action of the simulated spin Hamiltonian, $H_S$. The fidelity of the simulation is given by the probability of finding $| \psi_f \rangle$ after the real evolution:
\begin{eqnarray}
&&{\cal F}(\psi_i) = \langle  e^{-i H t/\hbar} \ \rho_i \otimes \rho_{ph} \ e^{i H t/\hbar} \rangle_{\psi_f} =  \label{fidelity} \\
&&\langle e^{\cal S} e^{-i (H_0 + H_E)t/\hbar} e^{-{\cal S}} 
\ \rho_i \otimes \rho_{ph} \ e^{\cal S} e^{i (H_0 + H_E)t/\hbar} e^{-{\cal S}} \rangle_{\psi_f} \nonumber ,
\end{eqnarray}
where $\rho_i = |\psi_i \rangle \langle \psi_i|$,and the vibrational modes are initially in a mixed thermal state ($\rho_{ph}$). 
$H_0 = H_v + H_S$, is the Hamiltonian without residual spin-phonon coupling.
In order to simplify the discussion, we define
$\eta = F \sqrt{\hbar/2 m \omega}/\hbar \omega$ ($\alpha$ is omitted), as the parameter that characterizes the displacement of the vibrational modes, and estimate the error, $E = 1- {\cal F}$, in a series in $\eta$.
If we use only one force in a given direction, then $H_E = 0$, and the error has its origin in the canonical transformation.
A straightforward expansion of (\ref{fidelity}) in ${\cal S}$, allows to estimate a correction $E \propto \eta^2 (1+ 2 \bar{n})$, where $\bar{n}$ is the mean phonon number. We estimate now the contribution from $H_E$. 
If two or three forces act on the ions, then $H_E$ gives an additional contribution that vanishes at zero temperature, $E \propto \bar{n}$.
We can avoid the need to cool the ions by imposing different trapping frequencies in each direction, so that only non-resonant terms appear in Eq. (\ref{error.heisenberg}). In this case, $E \propto J^2/(\omega_x - \omega_y)^2 = {\cal O}(\eta^4)$, and the main contribution to the error is again the one given by the canonical transformation. 

We illustrate how the effective spin-spin interaction works by presenting a numerical simulation for the case of two trapped ions.
We consider two lasers that push the ions with the same intensity in the radial ($x$, $y$) directions, so that an effective isotropic XY interaction ($J^x = J^y = J$) is simulated during a time $T = \pi/(8 J)$.
The evolution of the effective spins under such conditions corresponds to the application to the initial state of a $\sqrt{\textmd{SWAP}}$ gate
\cite{NielsenChuang}. 
The Hamiltonian includes the four vibrational modes in the radial directions, and the two effective spins that correspond to the internal states of the two ions. 
For the numerical solution, in the range or parameters considered here, it is enough to truncate the phonon Hilbert space by taking only the three lowest states in each mode. As a figure of merit, we have chosen the averaged fidelity, ${\cal F} = \int d \psi_i {\cal F} (\psi_i)$, that describes the averaged accuracy of the simulation upon integration over all the possible initial states.

\begin{figure}
  \centering
  \resizebox{1.6in}{!}{%
    \includegraphics{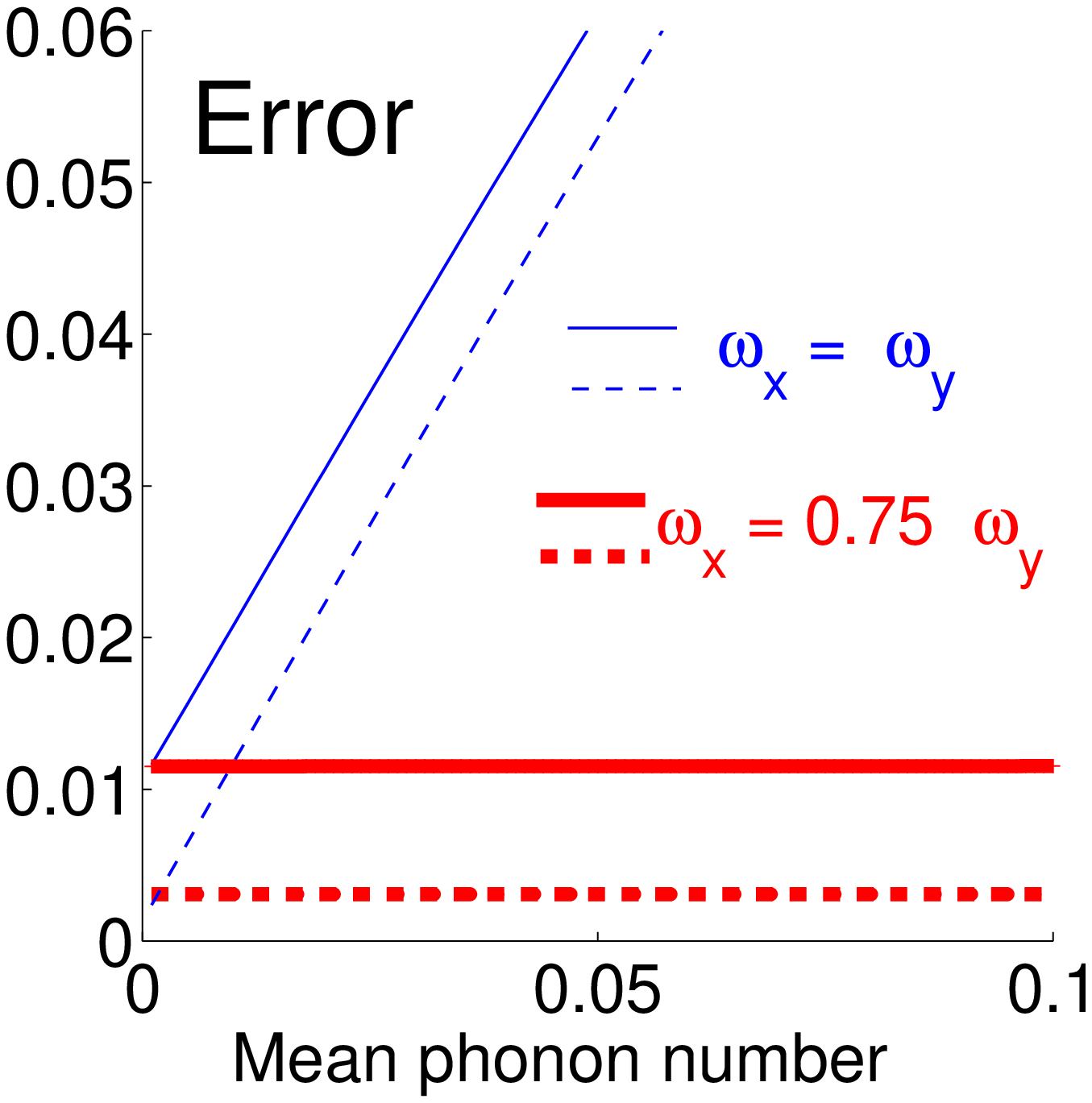}}
  \resizebox{1.6in}{!}{%
    \includegraphics{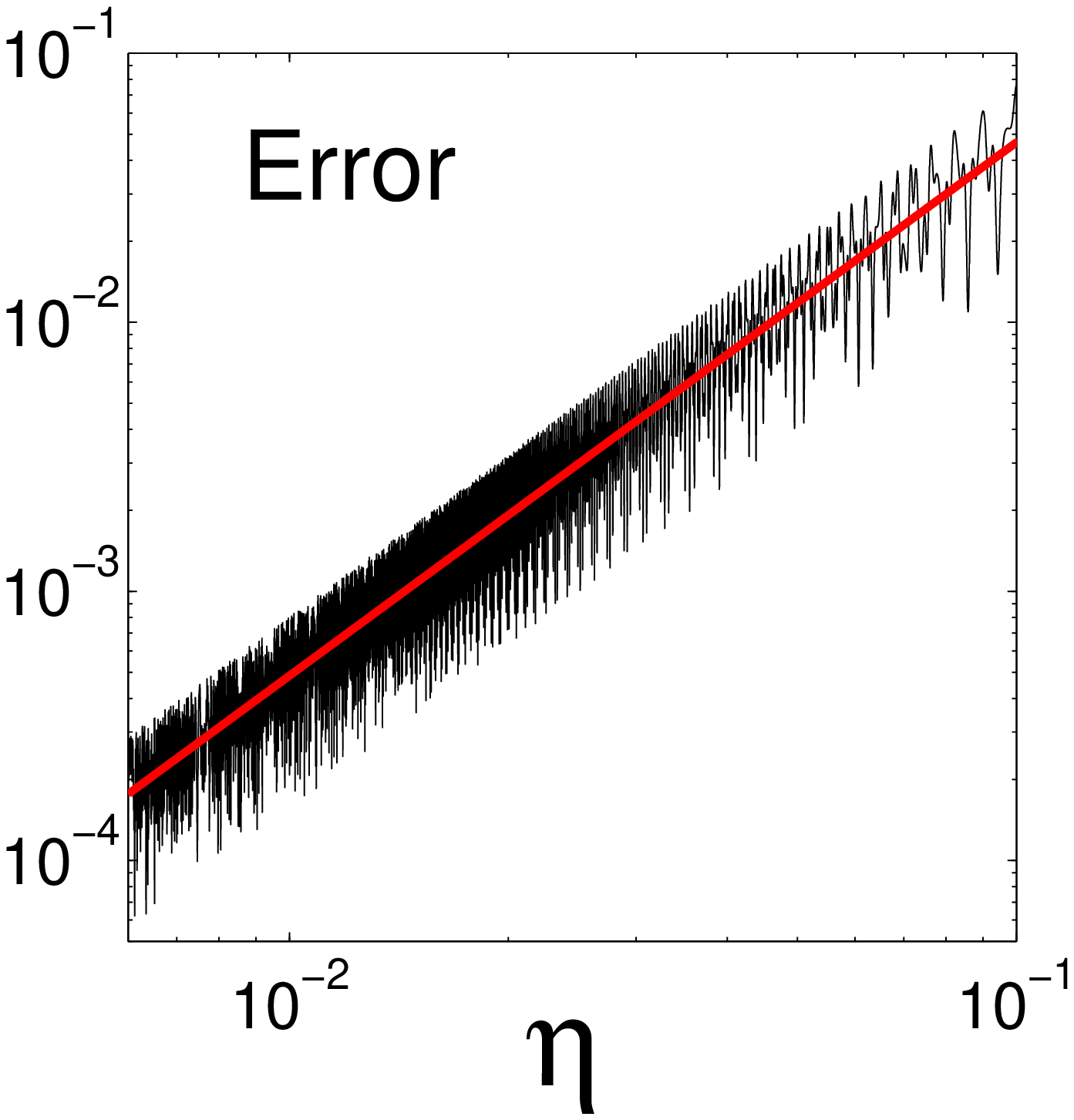}}
  \caption{(color online) Simulation of the XY model with two ions.
We have considered the case $\omega_x = 1.4 \omega_z$, and several values for $\omega_y$.
Left: Error of the simulation as a function of the mean phonon number $\bar{n}$, in a symmetric linear ion trap ($\omega_x = \omega_y$) (thin line), and in an asymmetric trap, $\omega_y = 0.75 \omega_x$ (thick line), with displacements $\eta = 0.05$ (continuous line), and $\eta = 0.02$ (dashed line). 
Right: Error of the quantum simulation as a function of the displacement of the phonons ($\eta$) for the asymmetric trap, and $\bar{n}=0.25$. The oscillations are due to the non-resonant contributions from each vibrational mode, and the averaged error satisfies the relation $E \propto \eta^2$ (straight line).}
  \label{fig3}
\end{figure}
The main results of our numerical calculation are illustrated in Fig. \ref{fig3}.
By introducing a small anisotropy ($\omega_x \neq \omega_y$) in the radial trapping frequencies, we get rid of the dependence of the error on the mean number of phonons, and the requirements for the cooling of the trapped ions are not so stringent (Fig. \ref{fig3} (left)). In Fig. \ref{fig3} (right) we show the dependence of the error on the displacement $\eta$, for the case of the asymmetric trap. With occupation numbers in the radial modes of $\bar{n} \approx 0.25$, and $\omega_z = 10$ MHz, our Eq. \ref{effective.interaction.2} allows us to calculate an interaction strength $J/\hbar \approx 10$ kHz, with error $E = 10^{-2}$. %

{\em Conclusions and Outlook.}
We have presented a method to simulate Ising, XY, and XYZ interactions between effective spins with trapped ions. Our scheme relies on the coupling of the internal states with the collective vibrational modes, and would lead to the observation of quantum phase transitions in ion traps. 
With some modifications, our proposal could also be used to simulate many other systems of great interest in Condensed Matter Physics. For example, the action of pushing forces on a single ion, could allow to realize a spin-boson model, with the bosonic bath being given by the vibrations of the ions. Different laser configurations than the one presented here, could also allow to simulate other particular spin hamiltonians. 

We acknowledge interesting discussions with D. Leibfried.
Work supported by EU IST projects, the DFG, and
the Kompetenz\-netz\-werk Quanten\-informations\-verarbeitung der
Bayerischen Staatsregierung.

\newpage

\end{document}